# Individual Polymer Chain Dynamics in an Entangled Polymeric Liquid Using a Stochastic Tube Model

Behrouz Behdani[1,2], Tong Mou[1], Cody Spratt[1], Slava Butcovich[1], Ryan Gettler[3], and Joontaek Park[1,3]*

[1]Department of Chemical and Biochemical Engineering, Missouri University of Science and Technology, Rolla, Missouri
[2]Department of Chemical and Biological Engineering, Vanderbilt University, Nashville, Tennessee
[3]Department of Biomedical, Biological and Chemical Engineering, University of Missouri, Columbia, Missouri

*Correspondence: parkjoon@missouri.edu

**Abstract**

This study focuses on comparing the individual polymer chain dynamics in an entangled polymeric liquid under different shear and extension rates. Polymer chains under various shear rates and extension rates were simulated using a stochastic-tube model [J. Rheol. 56: 1057 (2012)]. We developed a Matlab code to visualize and analyze the simulated configurations from the stochastic-tube model. We introduced new variables to determine how the extent of linearity changes with time for different shear rates, which is more useful than a typical end-to-end distance analysis. We identified whether the polymer chains undergo a "tumbling" motion (rotation with slight elongation not accompanying contraction) or "flipping" motion (rotation with elongation accompanying contraction). The simulation results indicate that the polymer chains exhibit a significant tendency to elongate at higher shear rates and occasionally experience flipping, while lower shear rates tend to exhibit very frequent tumbling. Furthermore, no rotations were observed under uniaxial extensional flows. These results help clarifying uncertainty of previously proposed polymer deformation mechanisms of the convective constraint release and the configuration-dependent friction coefficient.

## I.   INTRODUCTION

Polymeric liquids possess complex properties that significantly deviate from Newtonian fluids. Therefore, understanding polymer dynamics mechanisms in different flow conditions is important in modeling physical and rheological behaviors of polymeric fluids. The purpose of researching individual chain dynamics is to elucidate bulk-averaged rheological properties of polymeric liquids in a molecular level. Although laboratory methods such as light scattering and birefringence can be used to investigate configurational changes of polymers in flow, they are not able to elucidate the individual chain dynamics of a melt. Instead, the results will display bulk fluid properties instead of bulk rheological and microstructural properties of linear polymer liquids [Sefiddashti et al. (2015)].

To analyze polymer dynamic behavior for single, isolated polymer chains, computational models can be utilized. Molecular dynamics (MD) simulation is one such approach, but computational load for large scale polymer simulations is a hindrance [e.g. Kremer & Great (1990)]. To offset this burden, stochastic models have been developed to simulate the polymer network dynamics as bead-spring chains surrounded by approximated constraints instead of simulating all the neighboring polymer chains. Several of these models have proven to be effective at identifying distinguished behaviors in entangled polymers without having the complex computational analysis and loss in structural information associated with mathematical models and MD simulations. These include models approximating surrounding polymer chains as slip-links [e.g. Schieber et al. (2007)], mean-field constraining virtual tubes [e.g. Xu et al. (2006); Park et al. (2012)], and primitive chain networks [e.g. Yaoita et al. (2008)].

Rheological behaviors of entangled polymeric liquids have previously been successfully modeled using the 'tube' concept (Figure. 1) [Doi & Edwards (1968)], which approximates surrounding polymer chains as tube-like constraints. Numerous models have been proposed for improved predictions of nonlinear behaviors in fast flows or at high deformation rates [e.g. Mead et al. (1998; 2015)]. A typical approach is to assume a hypothetical mechanism and add a related correction term to a current model. Subsequent comparison of the predicted results with corresponding experimental data either confirms the model or suggest a new mechanism or mathematical term. However, these modeling procedures leave some ambiguity as to whether the hypothetical mechanism actually happens or a numerical correction term fortuitously improved the agreement. One such debate on polymer relaxation mechanisms under high deformation rates is the validity of the convective constraint release (CCR) mechanism, where tubes have a greater chance of elimination at higher deformation rates. The model predictions by Mead et al. (2015) showed that addition of a CCR term to their model improved the agreement with the experimental results obtained for shear viscosity at higher shear rates. However, steady extensional viscosity predictions were found to be better without CCR term. Another disputed term is the configuration-dependent friction coefficient (CDFC), also known as low-induced friction reduction [Ianniruberto (2015)], which describes how polymer chain friction decreases at high deformation rates due to aligned polymer structure [Park et al. (2012)].

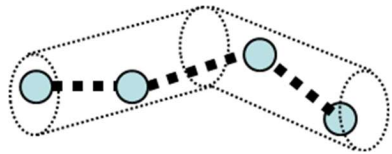

**Figure 1:** Schematic diagram of the concept of the stochastic tube model [Park et al. (2012)]: An individual polymer chain (bead-spring) is simulated as a typical bead-spring model which is constrained in a tube-like mean-field (cylinders in dashed lines), which follows the original tube model concept as it is [Doi & Edwards (1968)].

In spite of the effectiveness and simplicity of those stochastic models, additional computer programs to visualize and analyze the simulated structure of the polymer network are required. In this study, a Matlab code was developed to extract and manipulate a three-dimensional image from simulation results using the stochastic tube model developed by Park et al. (2012). Each image represents a single polymer chain at a particular point in time under a shear rate. Next, a movie file was made to represent the dynamic conformation of the polymer under a fixed shear rate over time. To determine whether the sigmoidal polymer behavior of viscosity versus shear rate can be correlated with rotational motions at the polymer level, a number of variables were introduced to quantify polymer rotation itself, including the linearity ratio, the static density ratio, dynamic density ratio and maximum distance in entangled polymers. Typical polymer structure analysis is based on the "end-to-end" vector of a polymer chain. We found that this analysis has difficulty in identifying whether the polymer chain is rotating due to swarming of the beads at low deformation rates.

 A recent study by Sefiddashti et al. (2015) performed nonequilibrium molecular dynamics (NEMD) simulations for a wide range of Weissenberg numbers in a polyethylene dense liquid and simultaneously performed Brownian dynamics (BD) simulations in dilute solution. As in many single polymer dynamics studies [e.g. Schroeder (2005); Monjezi et al. (2017ab)], they showed that polymer chains in entangled polymeric liquids are also "flipping", in which a polymer chain is elongated and then contracted while rotating under shear flows. The results were explained by using end-to-end distance and orientation angle verses time, which did not comprehensively define correlations between flipping rotations and proposed relaxation mechanisms or changes in the surrounding entangled structure. In this work, the newly introduced variable based on the polymer linearity is used to distinguish whether the rotation is "flipping" (elongation and contraction) or "tumbling" (only slight elongation without contraction). This study identifies those rotation mechanisms in different flow conditions to clarifies the existence of the polymer behaviors in fast flows.

## II.  Method

 The details of the description of the stochastic tube model can be found in References [Xu et al. (2006); Park et al. (2012)]. This study focuses on the visualization of the results from the stochastic model. The stochastic tube model simulation was performed using polystyrene as the sample polymer. The information required by the Fortran code includes the average molecular weight of the polymer, the polydispersity index, the solvent properties (Tricresyl phosphate as the solvent for this study), the relaxation time, and the shear rate being observed. It is noted that this input condition is based on the data set [Kahvand (1995)] used in Park et al. (2012). Fifty isolated polymer chains, each with unique equilibrium/starting conformations, were simulated to show the dynamic polymer behavior under various constant shear stresses. The shear rates observed in this study are $0.4\ s^{-1}$, $4.0\ s^{-1}$, and $40\ s^{-1}$. These shear rate corresponds to the intermediate (less than $\tau_R^{-1}$), high (larger than $\sim O(\tau_R^{-1})$), and very high shear rates (larger than $\sim O(10\tau_R^{-1})$) in terms of the Rouse time scale, $\tau_R$. These inputs were run with the stochastic-tube model Fortran code in order to generate data in the form of vector components, which indicate the spatial arrangement of each polymer chain at various times. The chains are made up of tubes and their respectively allocated beads, as demonstrated in Figure 1.

In order to display polymer chain conformation changes with time under a constant shear stress, a Matlab code was developed to extract the data, manipulate it, and produce three-dimensional plots showing the spatial conformation of the polymer chain in time intervals of 24 milliseconds or 120 milliseconds. The plot consists of the beads and the tubes for a single polymer chain. The calculation for the position vector (R) of the $i$th bead of the $j$th chain in the $k$th tube is as follows:

$$\vec{R}_{i,j,k} = S'_{i,j,k}\hat{u}_k + \sum_{\alpha=0}^{k-1} \vec{U}_\alpha + \vec{p}_{i,j} \tag{1}$$

where $S'_{i,j,k}$ is the scalar applied in the direction of the $k$th tube unit vector ($\hat{u}_k$) and $\vec{p}_{i,j}$ is the transverse translational position vector which determines how far a bead extends from the axis of the respective tube. Because the polymer beads are plotted from one end to the other, the tubes prior to the $k$th tube must be accounted for via vector addition. Hence, the summation of $\vec{U}_n$, a primitive path vector of $n$th tube, is present also. As such, $\vec{U}_0 = 0$, for there is no tube preceding the first.

Consequently, the tube end locations were determined by the following equations:

$$\vec{U}_{k,front} = \sum_{\alpha=0}^{n-1} \vec{U}_\alpha \tag{2}$$

$$\vec{U}_{k,back} = \sum_{\alpha=1}^{n-1} \vec{U}_\alpha \tag{3}$$

Obviously, $\vec{U}_{k,back}$ and $\vec{U}_{k+1,front}$ are equivalent. The tubes were drawn as a green line connecting the two points to illustrate the dependence of the beads on the tube locations. The plotted images were saved as jpeg files and converted to gif files in order to observe the polymer deformation in time at a constant shear rate.

In order to determine whether the sigmoidal polymer behavior of viscosity versus shear rate can be correlated with microscale tumbling action, a number of variables were introduced to quantify polymer tumbling. The linearity ratio ($\sigma_L$) is a dimensionless value existing between zero and one that expresses how linear a polymer chain conformation is at a particular point in time. It is calculated as shown in equation (4):

$$\sigma_L = \frac{|\vec{R}_n - \vec{R}_1|}{\sum_{\alpha=2}^{n}|\vec{R}_\alpha - \vec{R}_{\alpha-1}|} \tag{4}$$

where $n$ is the total number of beads in the polymer.

The denominator of equation (4) is the total length of the polymer chain, as demonstrated by the summation of the distance between every adjacent pair of beads along the polymer's length. The numerator of equation (4) is the distance between the first and last beads in the polymer. Hence, the linearity ratio would be unity if the angles between each bead were 180°.

The linearity ratio alone does not sufficiently convey whether the polymer is in a tumbling conformation, as a folded conformation is also likely to occur in high shear rates, whereby the two ends of the polymer may be close together without tumbling occurring. Therefore, another dimensionless value known as the density ratio ($\sigma_D$) is introduced.

The density ratio is calculated as shown in equation (5):

$$\sigma_D = \frac{\left(\frac{n}{V_a}\right)}{\left(\frac{n}{V_{d_{max}}}\right)} = \frac{d_{max}^3}{a^3} = \frac{\max\left(|\vec{R}_\alpha - \vec{R}_\beta|_{\alpha \neq \beta}\right)^3}{\left(\sum_{\alpha=2}^{n}|\vec{R}_\alpha - \vec{R}_{\alpha-1}|\right)^3} \qquad (5)$$

where $V_a$ and $V_{d_{max}}$ are reference volumes corresponding to a cube with a side length of $a$ and $d_{max}$, respectively. The distance, $a$, is defined as the total polymer length at a particular point in time. For the same point in time, $d_{max}$ is defined as the maximum distance between any two beads in a given polymer chain. In a fashion similar to the linearity ratio, the density ratio is a value between zero and one. A small value of $\sigma_D$ would imply that the polymer is very tightly packed; this signifies a tumbling motion under shear stress. Conversely, a large $\sigma_D$ would imply that the polymer is loosely packed; this signifies the lack of tumbling motion under shear stress.

Utilizing both the linearity ratio and the density ratio, one can select a range of values which qualify tumbling motion. As such, a tumbling conformation can be observed for polymer chains that exhibit a linearity ratio less than 0.5 and a density ratio less than 0.25. The percentage of time for which the polymer chains demonstrated tumbling over the total time of observation was calculated for each shear rate.

### III.   RESULTS

The figures shown below are examples of typical polymer chain behavior at each shear rate tested. The figures are presented in ascending order of shear rates for which the trial data were procured. In Figure 2 below, the spatial conformation for Chain 38 is shown as a snapshot at 15.072 seconds into the simulation for a shear rate of 0.4 s$^{-1}$ (an intermediate shear rate region).

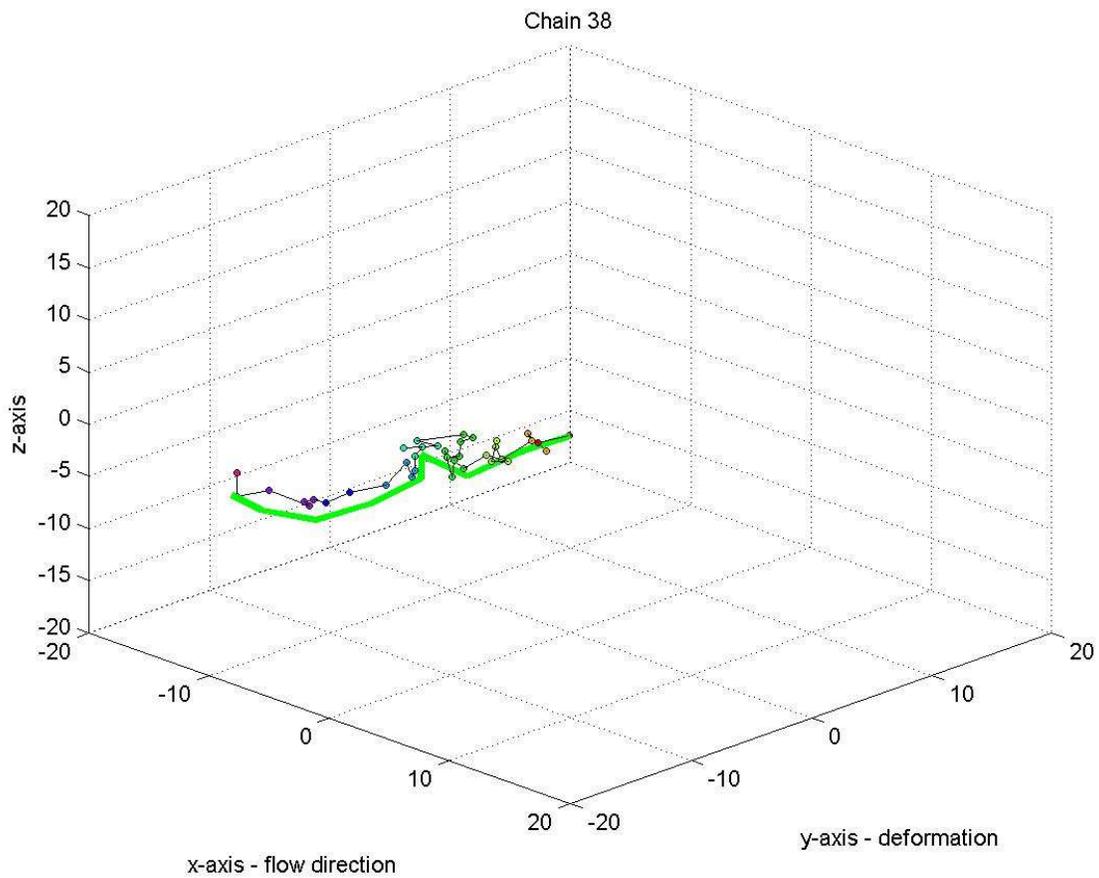

**Figure 2:** Chain 38 Plot; $\dot{\gamma}$ = 0.4 s$^{-1}$, $t$ = 15.072 s

Figure 2 shows Chain 38 in one of its least densely packed conformations, for which the linearity ratio and density ratio are both at near their maximum values for the observed time at a shear rate of 0.4 s$^{-1}$. A large degree of contortion can be observed. Such contortion lowers the values of the density ratio and the linearity ratio, hence increasing the likelihood for the criteria for tumbling to be met. Figure 3 shows the linearity ratio for Chain 38 for the time range observed in the simulation.

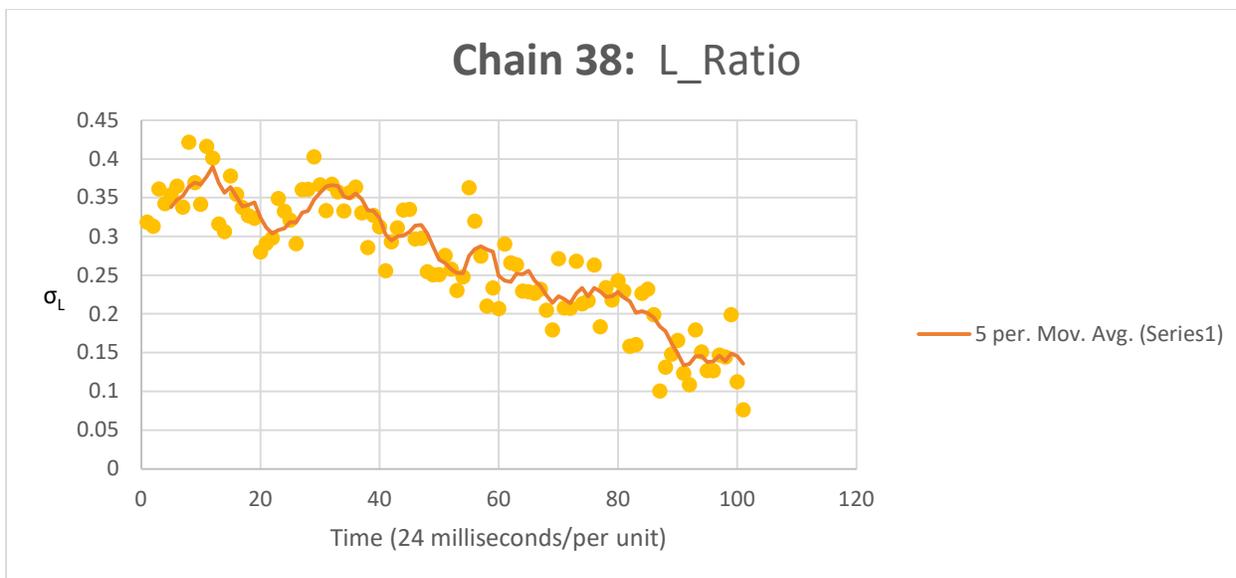

**Figure 3:** Chain 38 Linearity Ratio; $\dot{\gamma} = 0.4$ s$^{-1}$

    The linearity ratio for Chain 38 never reaches a value exceeding 0.5, satisfying the linearity ratio criterion for tumbling motion. Similarly, the density ratio values for Chain 38 displayed in Figure 4 never exceed 0.25, and therefore both criteria indicating tumbling motion are met for all times observed at a shear rate of 0.4s$^{-1}$.

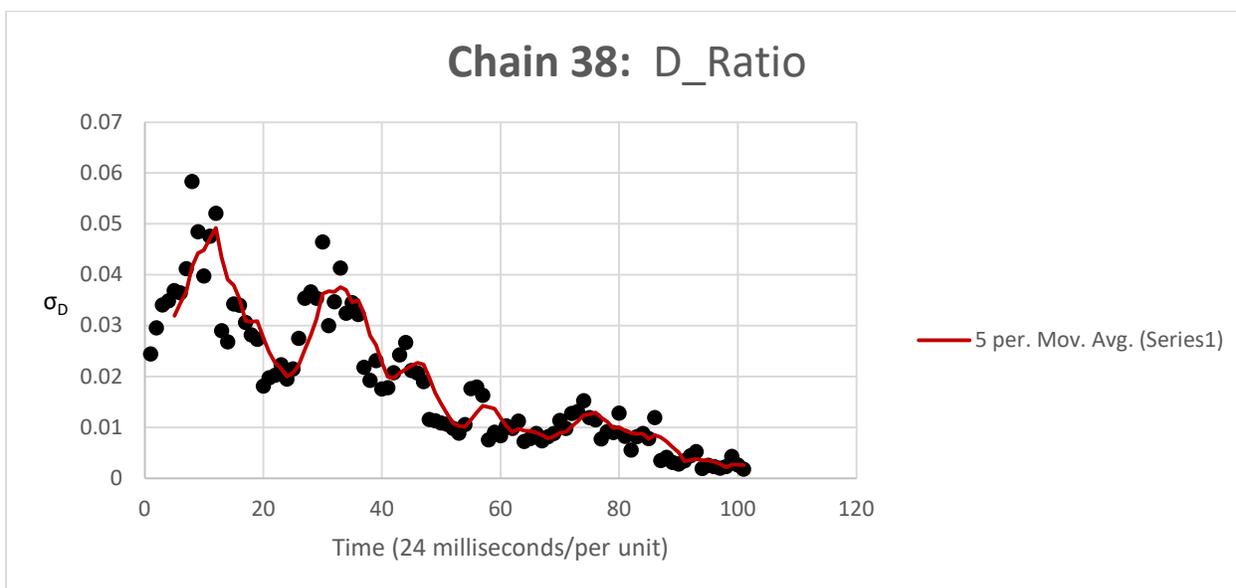

**Figure 4:** Chain 38 Density Ratio; $\dot{\gamma} = 0.4$ s$^{-1}$

    It was found that all fifty polymer chains displayed tumbling motion similar to Chain 38. Namely, for a shear rate of 0.4 s$^{-1}$, all polymer chains simulated with the stochastic-tube model under the experimental constraints employed by this study demonstrated a tumbling conformation for all times observed.

An order of magnitude increases in shear rate introduced flipping rotation not observed at 0.4 s$^{-1}$. In Figure 5 below, the spatial conformation for Chain 10 is shown as a snapshot at 26.420 seconds into the simulation for a shear rate of 4.0 s$^{-1}$ (high shear rate).

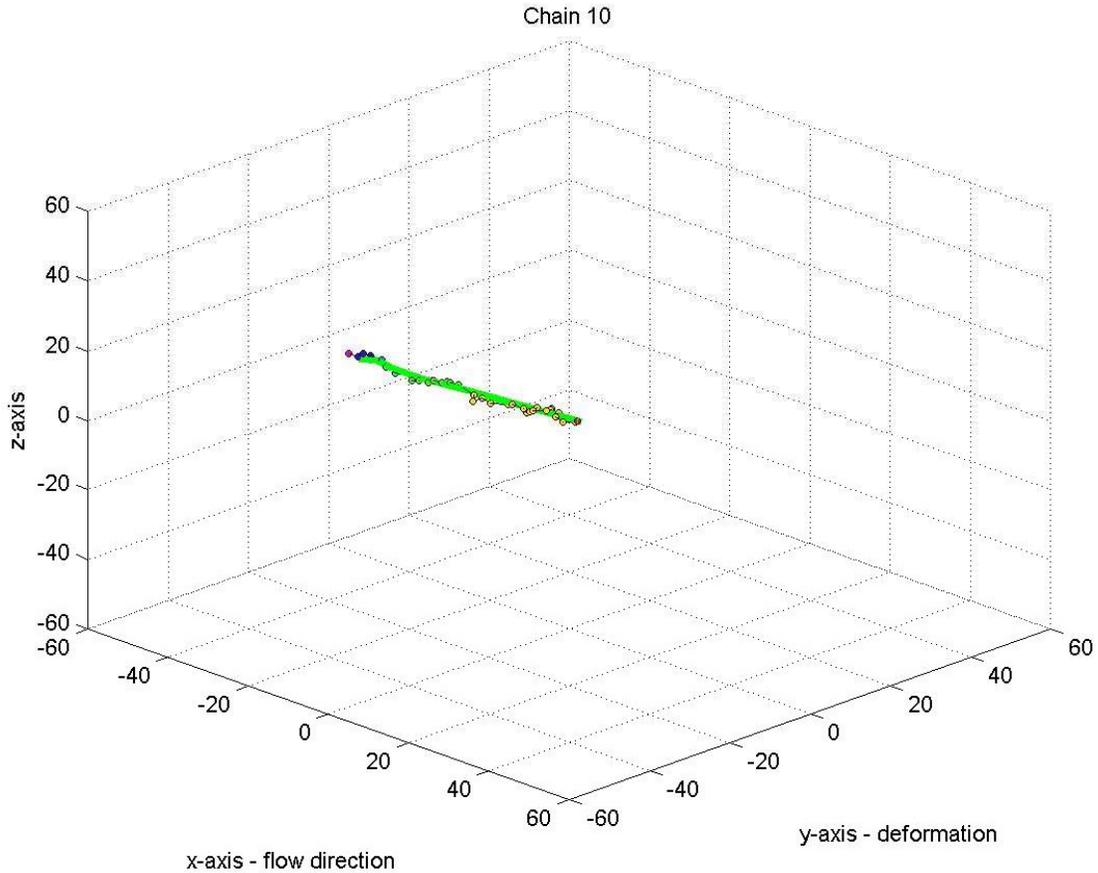

**Figure 5:** Chain 10 Plot; $\dot{\gamma}$ = 4.0 s$^{-1}$, $t$ = 26.420 s

Figure 5 displays Chain 10 in one of its least densely packed conformations. The linearity ratio and density ratio for this particular conformation are both at their maximum values for all times observed. Some contortion can be observed, but the conformation appears to be linear as a whole. The linearity and density ratio are approximately 0.687 and 0.262, respectively, hence, this conformation does not exhibit tumbling at this point in time. In sharp contrast, Figure 6 below displays Chain 10 in one of its most densely packed conformations.

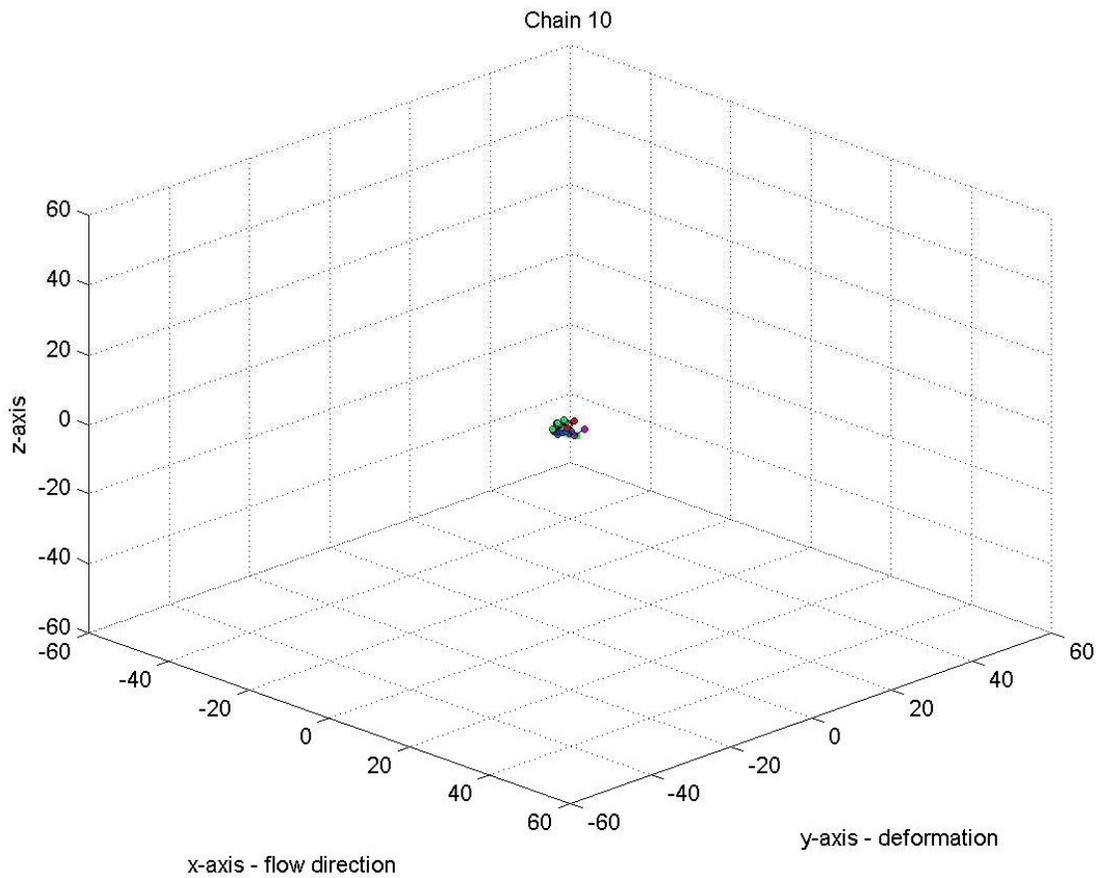

**Figure 6:** Chain 10 Plot; $\dot{\gamma}$ = 4.0 s$^{-1}$, $t$ = 28.700 s

As can be observed in Figure 6, this conformation is definitely exhibiting flipping motion. The linearity ratio is approximately 0.093 and the density ratio is approximately 0.002, which confirms the assertion that the polymer is tumbling at this point in time. Figure 6 below shows the linearity ratio for Chain 10 for the sample time range.

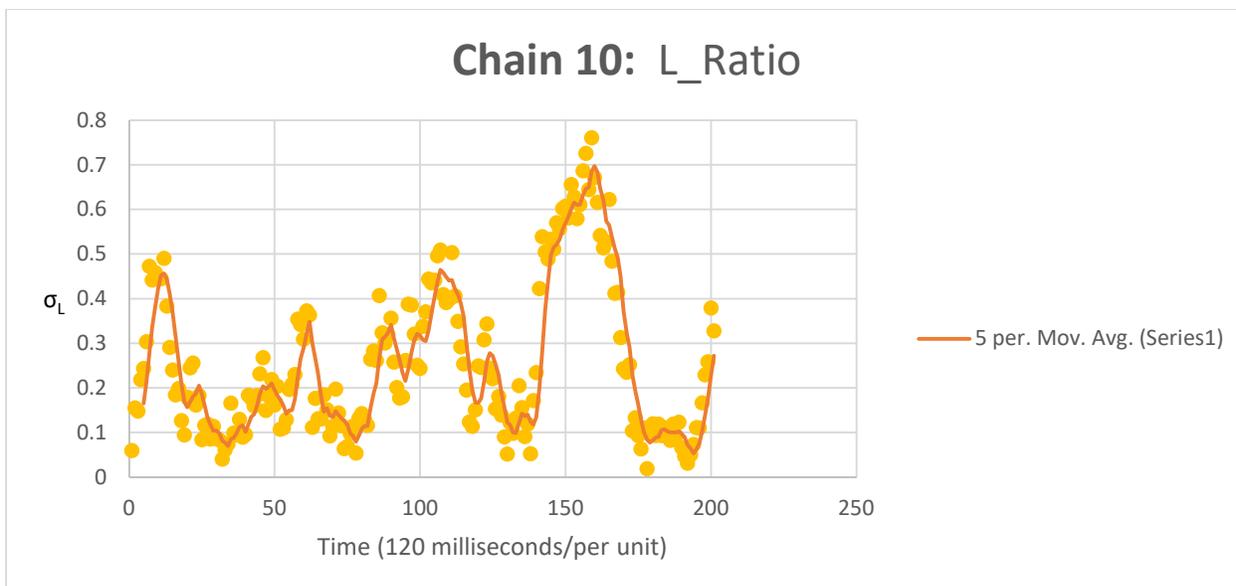

**Figure 7:** Chain 10 Linearity Ratio; $\dot{\gamma}$ = 4.0 s$^{-1}$

    By observing Figure 7, one notices that the linearity ratio value for Chain 10 fluctuates quite sporadically with time. This type of behavior is observed for all of the fifty chains at a shear rate of 4.0 s$^{-1}$. For the majority of the trial times, the linearity ratio does not exceed 0.5. Similarly, Figure 8 below demonstrates that the majority of trial times feature a density ratio below 0.25. As such, the percentage of the observed time for which Chain 10 maintained a tumbling conformation was 87.562%.

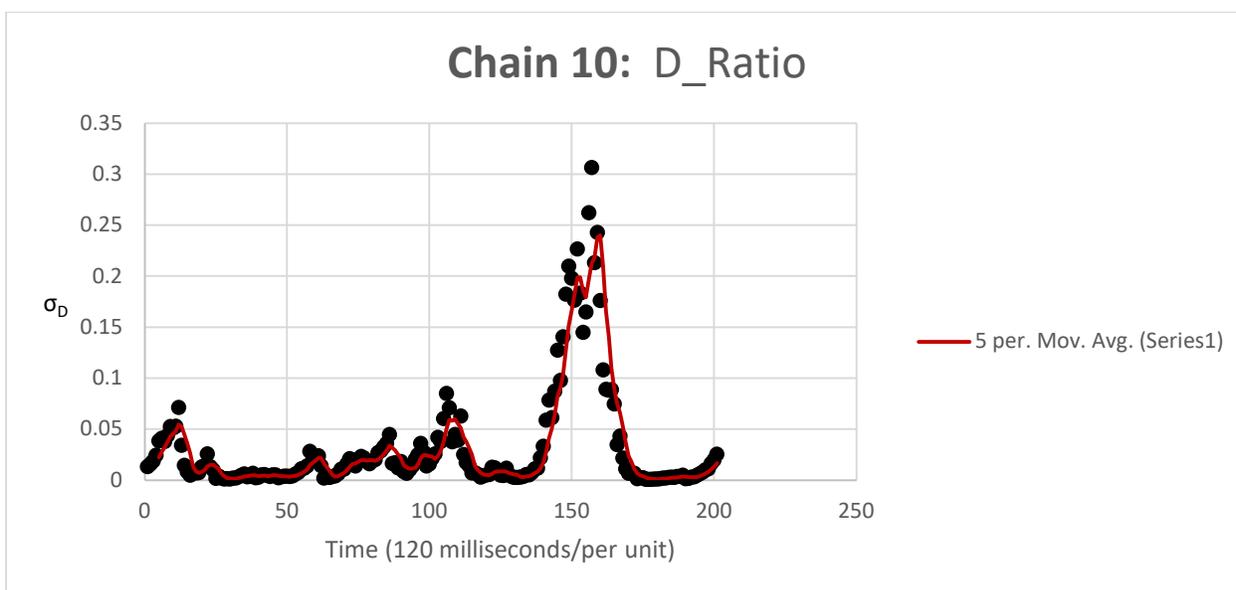

**Figure 8:** Chain 10 Density Ratio; $\dot{\gamma}$ = 4.0 s$^{-1}$

    This percentage is fairly typical of the polymer chains tested at a shear rate of 4.0 s$^{-1}$, for which the percentage of tumbling polymers was ≥ 80% throughout.

A further increase in shear rate decreases the probability of tumbling and flipping motion. In Figure 9 below, the spatial conformation for Chain 16 is shown as a snapshot at 7.700 seconds into the simulation for a shear rate of 40 s$^{-1}$.

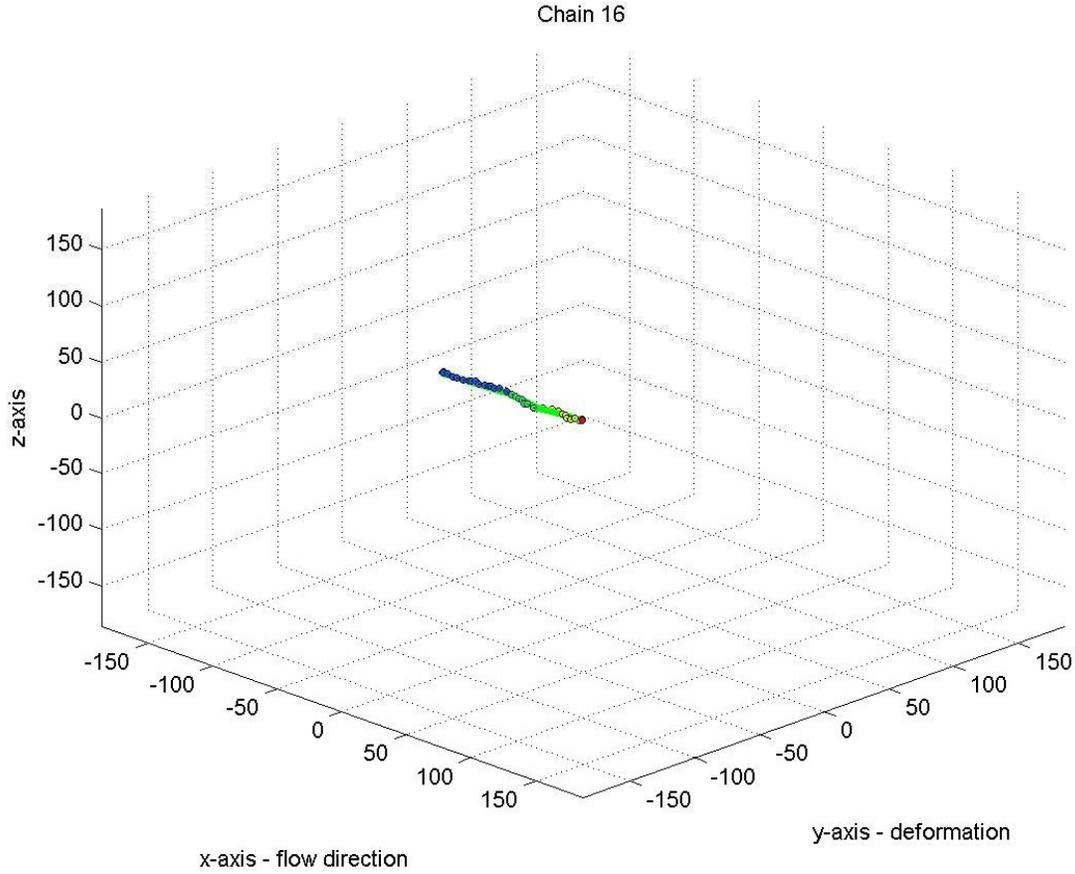

**Figure 9:** Chain 16 Plot; $\dot{\gamma}$ = 40 s$^{-1}$, $t$ = 7.700 s

Figure 9 demonstrates Chain 16 in a loosely packed conformation. This conformation is fairly typical for the first half of the trial for Chain 16. The second half of the trial bears a similar conformation (mostly linear, loosely packed), but in the reverse direction along the x-axis. This is shown in Figure 10.

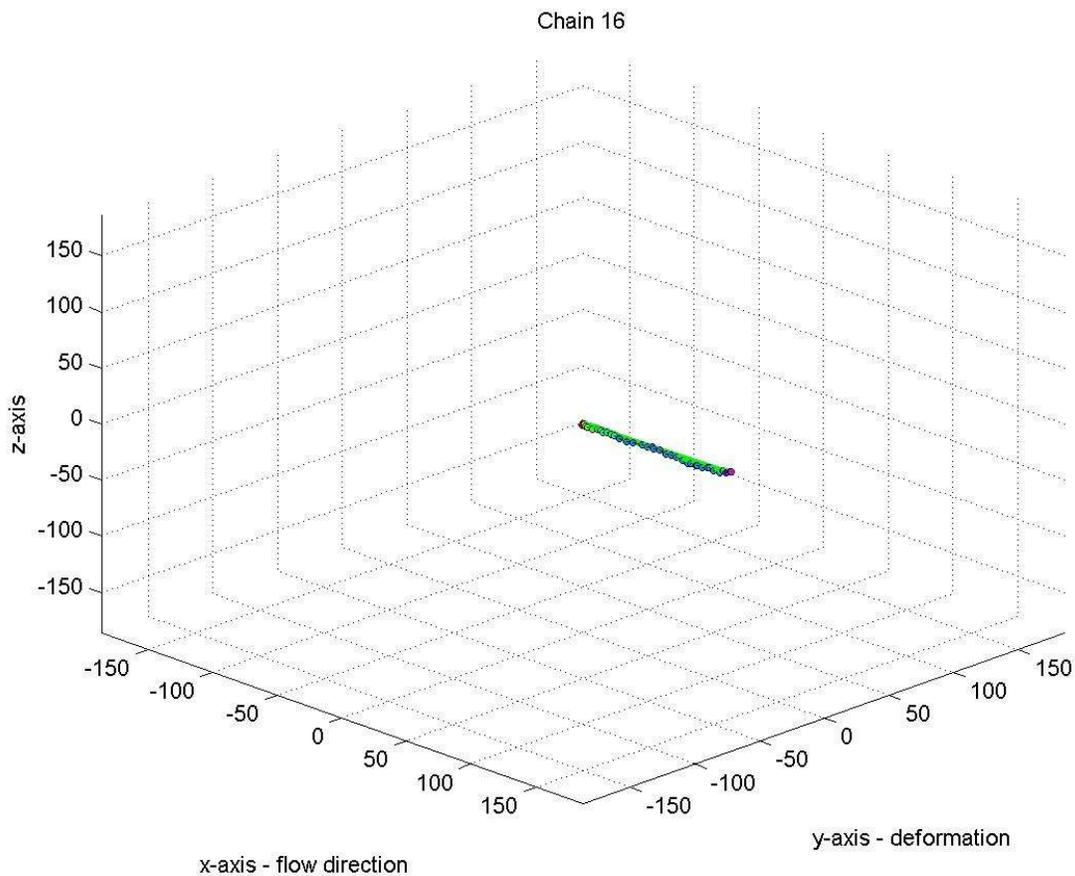

**Figure 10:** Chain 16 Plot; $\dot{\gamma}$ = 40 s$^{-1}$, $t$ = 9.548 s

    In order to obtain a conformation aligned in the opposite direction at a high shear rate, the polymer chain had to fold upon itself and "flip" in the spatial medium containing it. Figure 11 below shows the linearity ratio for Chain 16 at a shear rate of 40 s$^{-1}$ (a very high shear rate).

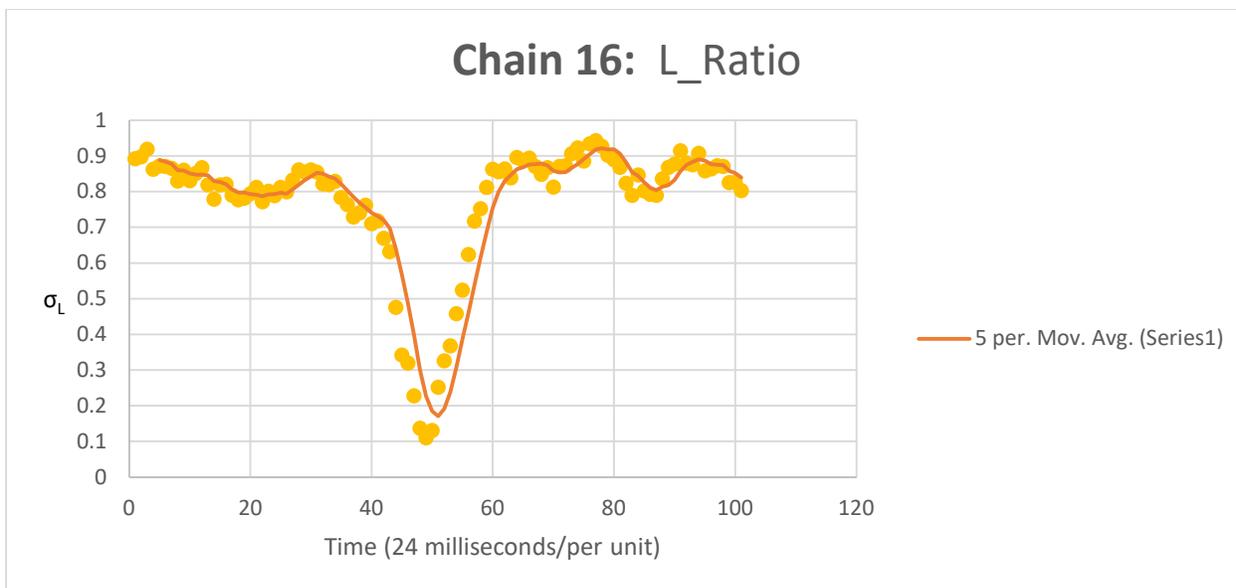

**Figure 11:** Chain 16 Linearity Ratio; $\dot{\gamma}$ = 40 s$^{-1}$

The conformational change that Chain 16 undergoes is reflected both in Figure 11 above and Figure 12 below. The low linearity ratio and density ratio values indicate chain contraction for a short timespan before reaching a state that exhibits no tumbling motion. Chain 16 experiences tumbling for 10.891% of the time observed, during which this conformational change takes place.

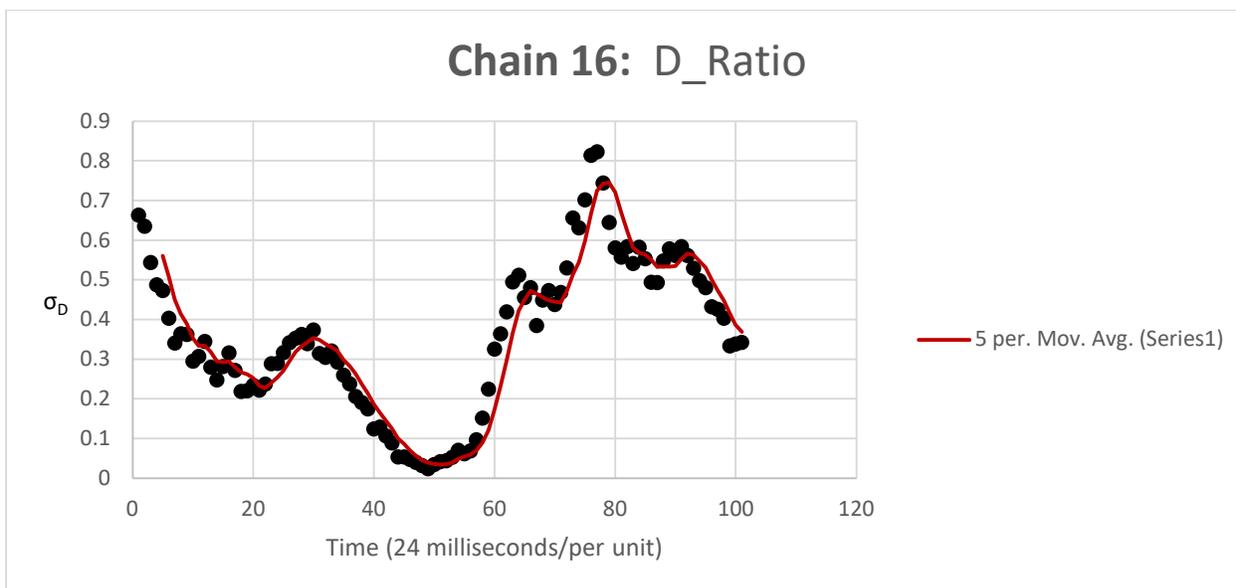

**Figure 12:** Chain 16 Linearity Ratio; $\dot{\gamma}$ = 40 s$^{-1}$

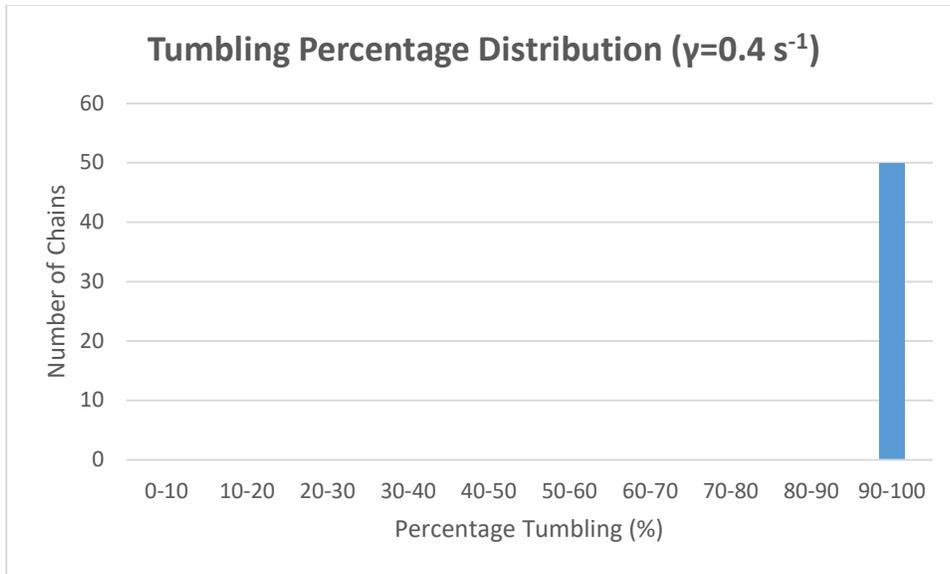

**Figure 13:** Number of Chains Tumbling for a Ranged Percent of Time; $\dot{\gamma}$ = 0.4 s$^{-1}$

As Figure 13 clearly shows, all fifty chains tested demonstrated that a large percentage of time was spent in tumbling motion. Also as mentioned before, all fifty chains maintained tumbling conformations for 100% of the time observed. Figure 14 demonstrates a similar tendency for the chains to maintain tumbling conformations for large time percentages.

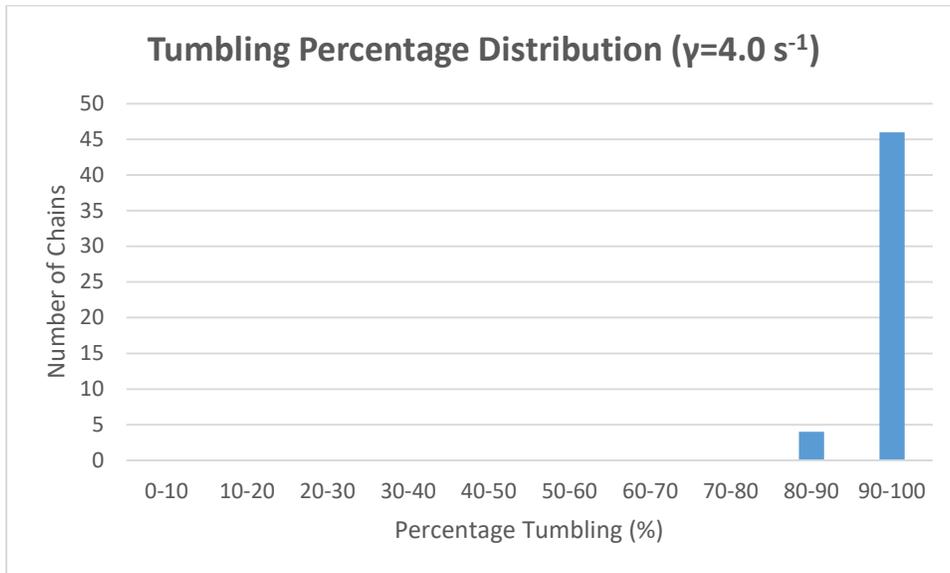

**Figure 14:** Number of Chains Tumbling for a Ranged Percent of Time; $\dot{\gamma}$ = 4.0 s$^{-1}$

The results in figure 14 imply that there is less tumbling at higher shear rates. This observation is further supported by Figure 15, which shows tumbling time percentages that not only vary much more for a shear rate of 40 s$^{-1}$, but display a bias for lower percentages.

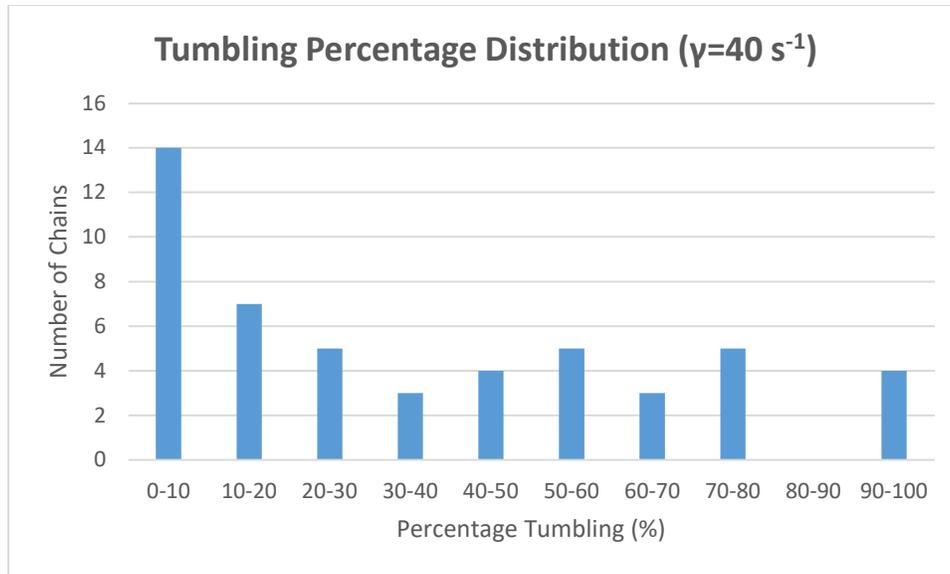

**Figure 15:** Number of Chains Tumbling for a Ranged Percent of Time; $\dot{\gamma}$ = 40 s$^{-1}$

## IV. DISCUSSION

The results from this study indicate that higher shear rates of more than 10 times of $\tau_R^{-1}$ result in a less frequent occurrence of flipping motion and less densely packed polymer conformations. Based on this study, it can be postulated that the elongated polymers under higher shear rates (fast flows) are in a conformation of least resistance to flow; such linear conformations are less subject to viscous forces. Conversely, a tumbling polymer (jumbled conformation) has a larger cross-sectional area in the direction of the flow and therefore would have a larger resistance to flow and be subjected to greater viscous forces in the sample medium. Hence, the general macroscopic sigmoidal polymer behavior of viscosity versus shear rate is indeed likely to be dependent on the spatial configuration of individual polymer chains at the molecular level. The tumbling rotation in low shear rates, which previously was not clearly identified, is resolved with the help of our newly introduced variables. Additionally, the mechanism of individual polymer chain rotations within the entangled polymer network are now explainable. Since the accompanying tubes are rotating along with the polymer chains, it follows that whole polymers networks are rotating together, which does not affect additional structural deformation.

The occurrence of flipping at a high shear of 4s$^{-1}$ ($>\tau_R^{-1}$) indicates that the CCR mechanism does occur at higher shear rates because the contraction of the chain releases tube constraints (tube elimination). However, at a very high shear rate of 40s$^{-1}$ ($>10\tau_R^{-1}$), the flipping frequency decreases and the polymer chain stays mostly in an elongated form. This conformation corresponds to the hypothetical structure for the reduction of the CDFC, where elongation of the polymer chains reduces friction. It can therefore be postulated that the CCR mechanism is switched to CDFC at a certain shear rate.

It is important to note that we also performed simulations under uniaxial elongational flows at different extension rates. The data is not shown because the polymer chain is only elongated without any rotations. This explains why the CCR mechanism made predictions of steady elongational viscosity worse in the previous work by Mead et al. (2015).

V.     CONCLUSION

We visualized and analyzed changes in polymer structure under shear and extensional flows, which were generated using a stochastic tube simulator. We introduced new variables, to identify tumbling and flipping rotations, subsequently improving upon the typical end-to-end analysis method. At low shear rates, polymer chains are predominantly experiencing tumbling motion. As shear rate increases, flipping rotation becomes dominant, which indicates the CCR mechanism taking place. However, as shear rate increases higher, flipping frequency decreases and CDFC prevails. No rotational motions were detected under extensional flow, which implies that CCR does not exist in extensional flow. Processing of polymers in high shear rate conditions has always been utilized in traditional polymer melt extrusion as well as recent advanced techniques, such as additive manufacturing [e.g. Mackay (2018); Behdani et al. (2020)]. This study and approach can provide valuable insight into the mechanisms of conformational changes in the polymer during processing.  Additionally, the results of this work can be applied toward improving current modeling of physical and rheological polymer behaviors.

VI.    NOMENCLATURE

| | |
|---|---|
| $\vec{R}$ | bead position vector |
| $\vec{U}$ | primitive path tube vector |
| $\vec{p}$ | transverse position vector |
| $S'$ | tensile adjustment scalar |
| $\sigma_L$ | polymer linearity ratio |
| $\sigma_D$ | polymer density ratio |
| $n$ | total number of beads |
| $a$ | total length of a polymer chain |
| $d_{max}$ | maximum distance between any two beads in a polymer chain |
| $V_a$ | cubic reference volume if entire polymer chain was linear, side length = $a$ |
| $V_{d_{max}}$ | cubic reference volume for polymer chain, side length = $d_{max}$ |

VII.   ACKNOWLEDGEMENTS

This work was partially supported by the University of Missouri Research Board and the OURE (Opportunities for Undergraduate Research) of Missouri University of Science and Technology.

## VIII. REFENERCES